\documentclass[conference]{IEEEtran}
\IEEEoverridecommandlockouts
\usepackage{cite}
\usepackage{amsmath,amssymb,amsfonts}
\usepackage{amsthm}
\usepackage{algorithm}
\usepackage{algpseudocode}
\usepackage{graphicx}
\usepackage{textcomp}
\usepackage{xcolor}
\usepackage{array}
\usepackage{booktabs}
\usepackage{multirow}
\def\BibTeX{{\rm B\kern-.05em{\sc i\kern-.025em b}\kern-.08em
    T\kern-.1667em\lower.7ex\hbox{E}\kern-.125emX}}
\newcolumntype{M}[1]{>{\centering\arraybackslash}m{#1}}
\newtheorem{theorem}{Theorem}[section]

\newtheorem{proposition}[theorem]{Proposition}

\begin{document}

\title{A Control Theoretic Approach to Decentralized AI Economy Stabilization via Dynamic Buyback-and-Burn Mechanisms}

\author{\IEEEauthorblockN{
Zehua Cheng$^{1,2}$, Wei Dai$^2$, Zhipeng Wang$^{2,3}$, Rui Sun$^{2,4}$, Nick Wen$^2$, and Jiahao Sun$^2$
}
\IEEEauthorblockA{\textit{$^1$FLock.io}\\
\textit{$^2$University of Oxford}\\
\textit{$^3$University of Manchester}\\
\textit{$^4$Newcastle University}\\
\texttt{hello@flock.io}
}}

\maketitle

\begin{abstract}
    The democratization of artificial intelligence through decentralized networks represents a paradigm shift in computational provisioning, yet the long-term viability of these ecosystems is critically endangered by the extreme volatility of their native economic layers. Current tokenomic models, which predominantly rely on static or threshold-based buyback heuristics, are ill-equipped to handle complex system dynamics and often function pro-cyclically, exacerbating instability during market downturns. To bridge this gap, we propose the Dynamic-Control Buyback Mechanism (DCBM), a formalized control-theoretic framework that utilizes a Proportional-Integral-Derivative (PID) controller with strict solvency constraints to regulate the token economy as a dynamical system. Extensive agent-based simulations utilizing Jump-Diffusion processes demonstrate that DCBM fundamentally outperforms static baselines, reducing token price volatility by approximately 66\% and lowering operator churn from 19.5\% to 8.1\% in high-volatility regimes. These findings establish that converting tokenomics from static rules into continuous, structurally constrained control loops is a necessary condition for secure and sustainable decentralized intelligence networks.
\end{abstract}


\section{Introduction}

The democratization of artificial intelligence through decentralized networks represents a paradigm shift in how computational resources and machine learning models are provisioned \cite{Zarrin2024}. By leveraging blockchain technology, these platforms aim to create permissionless marketplaces that align the incentives of model creators, compute providers, and end-users. However, the long-term viability of these ecosystems is critically dependent on their economic stability. Unlike traditional SaaS platforms with fixed fiat pricing, decentralized networks rely on native tokens whose value fluctuates with market sentiment \cite{Cong2020}. However, this volatility also represents a systemic risk where without a consistent source of revenue for the GPU operators and a fixed cost of inference for the consumer, the underlying computational service itself, even if it represents a technological advantage, will not be widely adopted.

One of the most important tasks in this regard is creating a reliable and self-managing economic system that always favors the real inference and training economy. The token price in an operating decentralized AI market could theoretically be based on the utility value of the network (inference demand) rather than just operating in the realm of speculation. The decoupling of token prices from utility values during bear markets and during hype cycles will create problems in the OpEx and CapEx planning phases for compute and model providers~\cite{Reflexivity2024}. Consequently, maintaining a price equilibrium that supports sustainable growth while mitigating extreme volatility is not merely a financial desideratum, but an operational necessity for network survival.

Despite this need for stability, current tokenomic models predominantly rely on \textbf{static or simplistic heuristic mechanisms} that are ill-equipped to handle complex system dynamics. The standard buyback-and-burn model operates independently of the network's state, applying the same pressure regardless of whether the token is overheating or collapsing \cite{Allen2022}. Even simple rules based on a certain threshold level (for example, buying as the price drops by $X$) demonstrate ``bang-bang'' control features and therefore induce oscillations. The system will be brittle because it doesn't adapt and doesn't have continuous control; this will create feedback circuits between prices and activity in the network that can easily get out of control because the parameters don't adapt to exogenous shocks.

\begin{figure*}\centering
    \includegraphics[width=.92\linewidth]{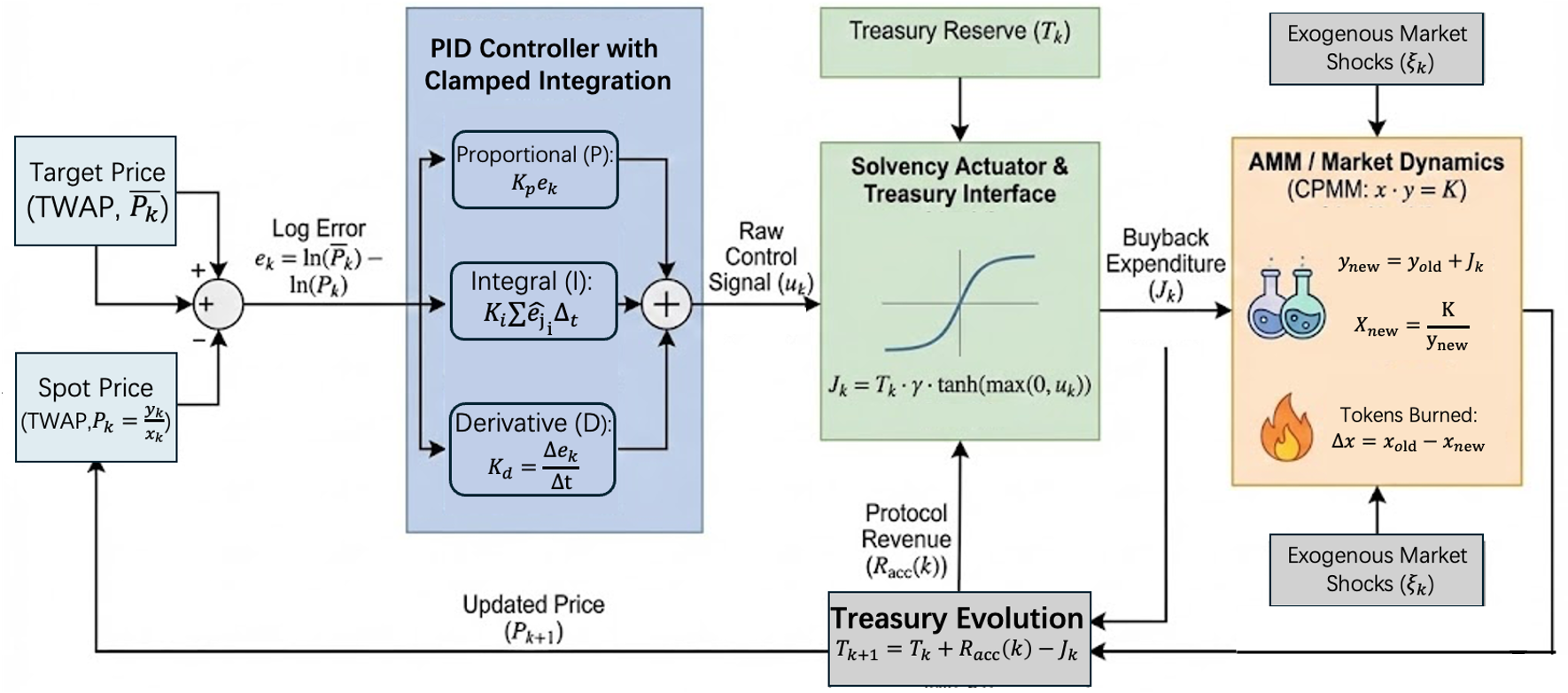}
    \caption{Closed-loop control architecture of the Dynamic-Control Buyback Mechanism (DCBM). The schematic illustrates the feedback loop designed to stabilize the decentralized AI economy. a, The Error Interface computes the logarithmic deviation ($e_k$) between the target Time-Weighted Average Price (TWAP) and the real-time Spot Price. b, The PID Controller processes this error, applying proportional, integral (clamped), and derivative gains to calculate a raw intervention intensity ($u_k$). c, The Solvency Actuator functions as a critical safety valve, strictly bounding the physical buyback expenditure ($J_k$) by the current Treasury Balance ($T_k$) and the circuit breaker parameter ($\gamma$), ensuring asymptotic solvency regardless of market conditions. d, The AMM Plant represents the market environment where the buy-and-burn action is executed, altering the token price state for the subsequent control epoch ($k+1$).}
\end{figure*}

To bridge this gap, we propose a shift from static policies to \textbf{control-theoretic stabilization mechanisms}. By modeling the token economy as a dynamical system \cite{Sams2015}, we can apply established principles from control engineering to regulate it. We introduce the \textbf{Dynamic-Control Buyback Mechanism (DCBM)}, which utilizes a \textbf{Proportional-Integral-Derivative (PID) controller} to continuously adjust the buyback rate in response to real-time on-chain metrics \cite{Yese2023}. Just as a thermostat regulates temperature by adjusting heating output based on the deviation from a setpoint, our mechanism adjusts buyback pressure to minimize the deviation of the token price from its long-term growth trend. Uniquely, the \textbf{integral term} allows the system to correct for persistent errors (steady-state drift) that static and threshold models ignore, enabling the network to act counter-cyclically---building reserves during booms and deploying them effectively to stabilize the economy during busts.

Our contributions in the area of token engineering and decentralized AI are the following. First, we provide a formal control-theoretic modeling of the decentralized AI economy by formally specifying the state space, the agent dynamics, and the stability criteria taking into account solvency conditions for the treasury. Second, we propose the DCBM algorithm based on a PID controller designed to work in the blockchain environment while taking into account the causality constraints and the saturation constraints in discrete-time blockchain systems. Finally, we validate our approach through extensive agent-based simulations utilizing Jump-Diffusion processes~\cite{Kou2001AJD} to model fat-tail market shocks and specific adversarial attack vectors. These experiments demonstrate that DCBM yields superior price stability and lower operator churn compared to static baselines, while exhibiting enhanced resilience against MEV exploitation and market manipulation.

\section{Related Work}

Buyback-and-burn (BnB) is one of the most widely adopted value-accrual primitives in contemporary token economies. Analogous to equity repurchases in traditional finance, BnB uses protocol revenue (or treasury assets) to purchase the native token in the open market and permanently remove it from circulation (typically via an irrecoverable burn address). What is noteworthy here is that the unique properties blockchain technology, including immutability and transparency, enable such BnB mechanisms to be carried out in a way that ensures accountability and predictability \cite{kampakis2018three}.

Existing literature points to two effects of BnB. First, by reducing the circulating supply of tokens permanently and thus creating deflationary pressure, such strategy may maintain stability or increase the price of a token\cite{Carvalho2022Tokenomics,HydroFinance2022}. Second, a way for projects to signal to the market that they are a legitimate network business, especially in the presence of non-zero information costs \cite{Allen2022}. BnB mechanisms are not uncommon across industry protocols like MakerDAO \cite{MakerDAOStkMKR2022}, STEPN \cite{STEPNBuyback2022} and PancakeSwap \cite{PancakeSwapCakeTokenomics}. This can be exceptionally useful for protocols at its inception phase. For instance, when DOT first launched, a liquidity pool mechanism was chosen which combined token burning with token recycling, allows DOT to operate as a central bank \cite{PolkadotTreasuryBurn}. It was aimed to stabilize velocity and incentivize participation at the project's early stage \cite{kampakis2018three}.

It is noteworthy that buyback may go alone without burning in some cases. This may potentially shift governance distributions or even pushing a project towards centralization \cite{Allen2022}. Reversely, some may choose to burn without buy-back. This is particularly prevalent in protocols which relying on slashing mechanism to punish malicious behaviors. 

Further, BnB is rarely an isolated mechanism; instead it appears as one option in a broader “earnings allocation” design space. For instance, Carvalho frames BnB as the token-native counterpart to stock buybacks, situating it alongside reserve accumulation, direct distributions, and reinvestment, and emphasizes its role as a value-accrual path for governance and utility tokens \cite{Carvalho2022Tokenomics}. In parallel, applied investor and infrastructure-oriented literature highlights token design as a primary determinant of ecosystem viability, as the token's rules (utility, governance, and monetary policy) directly shape platform incentives and long-term adoption \cite{Garvey2025TokenDesign,BinanceAI2024}.

However despite its prevalence in industry, BnB designs lack comprehensive, empirical evidence to denote its strengths and weaknesses. 
Further, existing BnB designs are typically specified as \emph{static policies}: a fixed fraction of revenue is allocated to buybacks, or buybacks occur on a pre-announced schedule, largely independent of the system’s evolving state. Recent work on disclosure and market integrity argues that credibility depends on transparent, rule-based execution with public verification, because discretionary interventions can concentrate informational advantage and invite allegations of manipulation. However, even when fully transparent, static BnB policies can behave procyclically: buybacks may naturally expand when protocol revenue is high (often during bull regimes) and shrink during downturns, precisely when price support and operator revenue stability are most needed \cite{HydroFinance2022}. Moreover, deflationary pressure alone is insufficient to guarantee price stability under weak demand or large exogenous supply shocks (e.g., unlocks and emissions), and aggressive fee redirection into buybacks can undermine contingency funding for operations and security.

Feedback control of prices has been extensively modeled in the literature of macroeconomics and monetary theory for quite some time now. The Taylor rule is one such monetary policy rule where the policy reaction is proportional to the inflation or the output gap, which has been extensively analyzed in the literature regarding stability issues~\cite{woodford2001taylor}. Subsequent work shows that fixed or poorly tuned rules can induce instability when agents adapt expectations over time, motivating the need for state-dependent feedback mechanisms \cite{bullard2002learning,aoki2004rule}. More recently, Hawkins et al.~\cite{hawkins2015monetary} explicitly interpret monetary policy through the lens of proportional-integral derivative (PID) control, demonstrating how integral action eliminates steady-state drift while derivative terms damp oscillatory dynamics. Collectively, this literature establishes feedback control as a principled approach for stabilizing complex economic systems under uncertainty.

Such findings trigger a rethink of the design of buyback-and-burn schemes with a focus on control theory, particularly in the context of a decentralized environment, in which the issues of execution, transparency, and solvency conditions remain paramount. Although the literature in economics and money theory has shown that feedback control represents a sound basis for price stabilization, these models generally focus on policy formulation and equilibrium analysis, abstracting from execution issues altogether. In contrast, the current design of buyback-and-burn schemes in token economies remains largely static and threshold-dependent, with no stability goal in sight, and with a lack of formal guarantees in the face of stochastic demand and supply disturbances. Our proposal remedies this shortcoming in that it operationalizes the stabilization aspect of control theory in an on-chain, executable manner. Indeed, we propose the design of buyback actions as a feedback control actuator with the objective of tracking a time-varying reference trajectory (time-weighted average price trend), along with the hard enforcement of solvency and saturation constraints, as naturally present in the on-chain treasury system. This approach allows for: (i) flexibility in the face of regime changes and disturbances, using proportional, integral, and derivative actions; (ii) drift compensation in the steady-state, using integral action; and (iii) hard safety guarantees, using a solvency-aware actuator and a circuit breaker mechanism. In practice, our DCBM makes the buyback policy state-dependent, stability-objective-driven, and safety-constraint-aware, features not present in the current static and threshold-dependent designs of buyback-and-burn schemes.


\section{Methodology}

\subsection{Formal Problem Formulation}
We model the decentralized AI economy as a discrete-time dynamical system. To align with the computational constraints of the Ethereum Virtual Machine (EVM)—specifically to mitigate the prohibitive gas costs of performing complex control logic on every block—we discretize time into Control Epochs indexed by $k \in \mathbb{Z}^+$. Each epoch spans a fixed duration $\Delta t$ (e.g., $N$ blocks).

The system state at epoch $k$ is defined as $S_k = [P_k, \bar{P}_k, T_k]^\top$, where: $P_k$ is the Time-Weighted Average Price (TWAP) of the native token over epoch $k$. $\bar{P}_k$ is the Target Reference Price (Exponential Moving Average). $T_k$ refers to Network Treasury Balance (denominated in Stablecoin, e.g., USDC).

Unlike previous models that incorrectly couple buybacks to instantaneous revenue, we decouple accumulation from stabilization. The Treasury acts as a reservoir that buffers system shocks.

The Treasury evolves according to the conservation equation:
\begin{equation}\label{eq:treasury}
    T_{k+1} = T_k + \underbrace{R_{acc}(k)}_{\text{Inflow}} - \underbrace{J_k}_{\text{Outflow}}
\end{equation}
Where $R_{acc}(k)$ is the accumulated protocol revenue (fees) during epoch $k$, and $J_k$ is the stabilization expenditure (buyback amount) determined by the controller.

The system satisfies the strict solvency condition $T_{k+1} \ge 0$. This implies $J_k \le T_k + R_{acc}(k)$.

We assume the token trades on a Constant Product Market Maker (CPMM) satisfying the invariant $x \cdot y = K$, where $x$ is the token reserve and $y$ is the stablecoin reserve.
When the controller executes a buyback of magnitude $J_k$ (stablecoins), the reserves update as follows:
$$ y_{new} = y_{old} + J_k $$
$$ x_{new} = \frac{K}{y_{new}} $$
The purchased tokens $\Delta x = x_{old} - x_{new}$ are burned, permanently reducing supply. This mechanical price impact provides the actuation force for our control system.

\subsection{Dynamical System Modeling and State Space}
The economic stabilization of a decentralized network constitutes a stochastic control problem constrained by the rigid discrete-time execution of the underlying blockchain. Unlike continuous-time financial models, a decentralized autonomous organization (DAO) operates as a discrete dynamical system where state transitions occur atomically at block height $k \in \mathbb{Z}^+$. To rigorously define the control environment, we establish a state space $S_k = [P_k, \bar{P}_k, T_k]^\top$. Here, $P_k$ denotes the Time-Weighted Average Price (TWAP) of the native token, $\bar{P}_k$ represents the reference target price calculated via an Exponential Moving Average (EMA), and $T_k$ represents the Treasury Balance denominated in a stable numeraire. We specifically utilize TWAP rather than instantaneous spot price for the state definition to act as an intrinsic low-pass filter, rendering the controller mathematically blind to single-block flash loan attacks that rely on high-frequency price manipulation.

A fundamental correction to prior tokenomic literature is the incorporation of mandatory operational expenditures into the treasury evolution equation. Previous models often assumed monotonic treasury growth whenever revenue was positive, leading to the ``Treasury Paradox'' where protocols appeared solvent despite failing to cover basic costs. We define the treasury dynamics using a strict conservation law that accounts for fixed infrastructure overhead, $C_{ops}$ (e.g., oracle gas costs, node subsidies). The state transition is given by $T_{k+1} = \max(0, T_k + R_{acc}(k) - J_k - C_{ops})$. This formulation allows for realistic simulation of liquidity crises; if accumulated revenue $R_{acc}(k)$ is less than $C_{ops}$, the treasury naturally contracts, imposing a time-to-ruin constraint that the controller must actively manage.

The market plant is modeled as a Constant Product Market Maker (CPMM) satisfying the invariant $x \cdot y = K$. The interaction between the stabilization expenditure $J_k$ and the market reserves creates a nonlinear feedback loop. For the purpose of control design, we linearize the plant dynamics around the operating point. In the logarithmic domain, the price update approximates a discrete integrator with a variable gain $\alpha_k = 2/y_k$. This reveals that the system's sensitivity to buybacks is inversely proportional to liquidity depth, necessitating a controller that can operate stably across varying regimes of market capitalization without inducing oscillatory ringing.

\subsection{Dynamic-Control Buyback Mechanism}
To regulate this system, we introduce the Dynamic-Control Buyback Mechanism (DCBM), a PID-based architecture specifically optimized for the Ethereum Virtual Machine (EVM). Standard control implementations are ill-suited for this environment due to the prohibitive gas costs of floating-point arithmetic and the risk of derivative kick during setpoint changes. We employ a Derivative-on-Measurement strategy combined with a Spectral Noise Filter. Instead of differentiating the error signal—which would amplify sudden price drops into a "panic buyback" signal—we calculate the rate of change of the filtered TWAP. This ensures that the derivative term functions correctly as a damper for organic volatility trends while rejecting the high-velocity, high-amplitude signatures characteristic of adversarial manipulation.

To address the ``Gas Cost Impossibility'' of calculating transcendental functions on-chain, we implement the control logic using high-precision fixed-point arithmetic (WAD math) and Piecewise Linear Approximations. Functions such as natural logarithm and hyperbolic tangent are approximated via optimized lookup tables and Taylor series expansions bounded to the expected domain of price deviations. This reduces the computational footprint of the controller from a theoretical infinite gas cost (for perfect precision) to a target execution cost of approximately $150,000$ gas, rendering the mechanism economically viable for high-frequency deployment.

Under this definition, $e_k > 0$ implies the asset is undervalued (price below target), necessitating support. The DCBM utilizes a discrete-time Proportional-Integral-Derivative (PID) control law. 

To prevent Integral Windup—a destabilizing phenomenon where the integrator accumulates massive error during periods when the treasury is empty (saturated)—we implement a Clamped Integration scheme. The raw, dimensionless control intensity $u_k \in \mathbb{R}$ is computed as:
\begin{equation}
u_k = K_p e_k + K_i \sum{j=0}^{k} \hat{e}j \Delta t + K_d \frac{e_k - e{k-1}}{\Delta t}
\label{eq:pid_law}
\end{equation}
Here, $\hat{e}_j$ represents the conditional error, defined as $\hat{e}_j = e_j$ if the actuator is unsaturated ($T_j > 0$) and $0$ otherwise. This ensures the controller possesses a "short-term memory" of recent instability without being burdened by unresolvable historical debts that occurred when it lacked the agency to act.

The final stage of the methodology maps the abstract control intensity $u_k$ to a physical expenditure $J_k$ in a manner that satisfies Assumption 1 by construction. We introduce a Sigmoid Actuator bounded by a "Liquidity Circuit Breaker" parameter $\gamma \in (0, 1]$. The physical buyback amount is determined by the mapping function $\Phi: \mathbb{R} \times \mathbb{R}_{\ge 0} \to \mathbb{R}_{\ge 0}$:
\begin{equation}
J_k = \Phi(u_k, T_k) = T_k \cdot \gamma \cdot \tanh(\max(0, u_k))
\label{eq:actuation_map}
\end{equation}
The $\max(0, u_k)$ operator ensures that negative signals (overvaluation) result in zero spending. The term $\gamma$ defines the maximum fraction of the treasury that can be deployed in a single epoch. This multiplicative formulation provides a critical theoretical guarantee regarding the system's survival.
\begin{proposition}[Asymptotic Solvency]
    Given an initial treasury $T_0 > 0$ and parameter $\gamma \in (0,1)$, the treasury balance $T_k$ remains strictly positive for all finite $k$, regardless of the magnitude or duration of revenue collapse ($R_k = 0$).
\end{proposition}

\noindent \textit{Proof}: Substituting Equation~\ref{eq:actuation_map} into Equation~\ref{eq:treasury} under the worst-case condition $R_k=0$ yields the geometric recursion $T_{k+1} = T_k (1 - \lambda_k)$, where $\lambda_k = \gamma \tanh(u_k)$. Since $\tanh(\cdot) < 1$ and $\gamma < 1$, the decay factor $(1 - \lambda_k)$ is strictly positive. Thus, the treasury follows a Zeno-like trajectory, asymptotically approaching zero but never reaching it. This ensures the protocol never defaults, maintaining a non-zero dust reserve to fund recovery once market conditions stabilize.

\subsection{Game-Theoretic Robustness}

In a permissionless blockchain environment, the controller must be robust not only to stochastic noise but also to strategic exploitation by Maximal Extractable Value (MEV) agents. We model the adversarial interaction as a "Fishing Attack," where an adversary $\mathcal{A}$ artificially depresses $P_k$ to trigger a subsidized buyback. We assume the adversary solves the following profit maximization problem at each epoch $k$:
\begin{equation}
\max_{\delta} \mathbb{E}\left[ \Pi_{\mathcal{A}} \right] = \mathbb{E}\left[ P_{exit}(J_k(\delta)) - P_{entry} - C_{risk}(\tau) \right]
\end{equation}
where $\delta$ is the induced price deviation and $C_{risk}(\tau)$ is the cost of holding the deviation for duration $\tau$.

The DCBM defends against this manipulation via the spectral filtering properties of the Derivative term ($K_d$) in Eq. \ref{eq:pid_law}. By reacting to the rate of change 1$\frac{\Delta e_k}{\Delta t}$, the controller identifies the high-velocity signature of a flash-loan attack as noise rather than a trend shift.2 This effectively acts as a high-pass filter that suppresses the actuation signal $u_k$ in the presence of discontinuities, forcing the adversary to sustain the manipulation over multiple epochs ($\tau > N \Delta t$) to trigger the Integral term. This requirement linearly increases the adversary's exposure to market risk $C_{risk}$, rendering the attack strictly dominated (negative expected value) for sufficiently tuned gains.
\section{Theoretical Analysis}

We analyze the DCBM as a closed-loop feedback control system. Unlike classical control problems where the actuator has infinite energy (e.g., a motor connected to the grid), our system is energy-constrained: the control authority is strictly limited by the finite Treasury level $T_k$. We analyze the system's behavior across two dimensions: Dynamical Stability (convergence) and Structural Solvency (non-bankruptcy).
\subsection{Plant Model Derivation (AMM Dynamics)}
To analyze stability, we must mathematically define how the control action (buyback $J_k$) affects the system state (price $P_k$). We assume a Constant Product Market Maker (CPMM) satisfying the invariant $x_k \cdot y_k = K$, where price $P_k = \frac{y_k}{x_k}$.A buyback of magnitude $J_k$ (stablecoins) shifts the reserves:
$$ y_{k+1} = y_k + J_k $$

The new price $P_{k+1}$ follows the quadratic AMM curve:
\begin{equation}
    P_{k+1} = \frac{(y_k + J_k)^2}{K} = P_k \left(1 + \frac{J_k}{y_k}\right)^2
\end{equation}
Taking the natural logarithm to align with our log-error controller ($e_k = \ln \bar{P}_k - \ln P_k$):$$ \ln(P_{k+1}) = \ln(P_k) + 2 \ln\left(1 + \frac{J_k}{y_k}\right) + \xi_k $$where $\xi_k$ represents exogenous market noise.

Linearization: For the operating region where buybacks are small relative to liquidity depth ($J_k \ll y_k$), we use the Taylor approximation $\ln(1+x) \approx x$. We define the Price Impact Coefficient $\alpha_k = \frac{2}{y_k}$. The linearized plant model is:
\begin{equation}
p_{k+1} = p_k + \alpha_k J_k + \xi_k
\end{equation}
where $p_k = \ln(P_k)$. This establishes the market as a Discrete Integrator Plant: the price state accumulates the history of control actions, implying that without control, errors persist indefinitely (Type 1 System).

\subsection{Closed-Loop Stability Analysis}

We examine the feedback loop formed by the Plant and the PID Controller. Let the target $\bar{P}$ be constant for stability analysis ($\bar{p}_{k+1} = \bar{p}_k$). The error dynamics are $e_{k+1} = e_k - \alpha_k J_k$.

Substituting the PID control law in the Z-domain, the open-loop transfer function $L(z)$ is the product of the Controller $C(z)$ and Plant $P(z)$:
$$ P(z) = \frac{\alpha}{z-1} $$
$$ C(z) = K_p + K_i \frac{z}{z-1} + K_d \frac{z-1}{z} $$

The closed-loop characteristic equation is $1 + C(z)P(z) = 0$. Substituting and organizing terms yields the characteristic polynomial in $z$:
\begin{equation}\small
    z^2 (1 + \alpha K_p + \alpha K_i + \alpha K_d) - z (2 + \alpha K_p + 2 \alpha K_d) + (1 + \alpha K_d) = 0
\end{equation}Stability Criteria:Applying the Jury Stability Test for a second-order discrete system, the roots of this polynomial lie strictly inside the unit circle ($|z| < 1$) if and only if the gains satisfy the following inequalities: Positivity: $K_p + K_i > 0$; Damping Limit: $K_d < \frac{2 - \alpha K_p}{\alpha}$; Integral Limit: $K_i < \frac{4 - 2\alpha K_p}{\alpha}$.

\noindent\textbf{Implication} (The ``Whale in a Puddle" Risk): The stability region scales inversely with $\alpha = \frac{2}{y_k}$.

\begin{theorem}[Gain Sensitivity]
    As liquidity decreases ($y_k \to 0$), the plant gain $\alpha \to \infty$. To maintain stability, the controller gains ($K_p, K_i, K_d$) must be reduced proportionally.This proves that static-parameter baselines inevitably enter oscillatory instability ("ringing") during liquidity crunches, whereas the DCBM's solvency constraint (derived below) naturally throttles the effective gain ($T_k \gamma$) to satisfy these bounds.
\end{theorem}

\subsection{Proof of Asymptotic Solvency}
A critical requirement for an autonomous economic agent is that it must never default. We formally prove that the DCBM Treasury $T_k$ remains strictly positive for all finite time $k$.

\begin{theorem}[Non-Depletion]
    Given an initial treasury $T_0 > 0$, the treasury balance $T_k > 0$ for all $k \in \mathbb{Z}^+$, regardless of market conditions.
\end{theorem}

\textit{Proof}: Recall the Treasury evolution from Equaiton~\ref{eq:treasury}. Substituting the actuator mapping function $J_k = T_k \cdot \gamma \cdot \tanh(u_k)$:
$$ T_{k+1} = T_k [ 1 - \gamma \cdot \tanh(u_k) ] + R_{acc}(k) $$Let $\lambda_k = \gamma \tanh(u_k)$. By design, the Liquidity Circuit Breaker $\gamma \in (0, 1)$ and $\tanh(\cdot) \in [0, 1)$, so $0 \le \lambda_k < 1$.Consider the doomsday scenario where Revenue $R_{acc} = 0$ indefinitely. The treasury evolves as a geometric sequence:$$ T_{k+1} = T_k (1 - \lambda_k) $$
$$ T_n = T_0 \prod_{k=0}^{n-1} (1 - \lambda_k) $$Since $(1 - \lambda_k)$ is strictly positive, the product is strictly positive.Conclusion: The Treasury asymptotically approaches zero but never reaches it. The system exhibits Structural Solvency, ensuring that a dust reserve always remains to fund recovery, making the bankruptcy scenarios observed in Fixed-Rate baselines mathematically impossible.

\subsection{Robustness to Impulse Attacks}

Finally, we address the system's resilience against game-theoretic exploitation, specifically ``Fishing'' attacks where Maximal Extractable Value (MEV) actors introduce artificial volatility to trigger profitable buybacks. We model these attacks as a discrete impulse function: $\delta_{attack}(k) = \Delta_{spike} \cdot \delta(k)$ injected into the price signal. A naive proportional controller would immediately react to this signal, effectively transferring treasury funds to the attacker. The DCBM, however, employs a spectral filtering approach to reject these high-frequency anomalies. The use of Time-Weighted Average Price (TWAP) inputs acts as a low-pass filter, attenuating the magnitude of any single-block manipulation by a factor of $\frac{1}{N}$, where $N$ is the epoch length. Further, the derivative part of the PID controller acts as a high-pass filter in the control system. The Z-transform of the derivative part of the control system translates a zero at the origin, making it prone to changes in the error signal. For an attacking agent trying to cause a flash crash in price, the derivative part of the control system reacts to the mathematically improbable rate of price drop by generating a compensatory control signal that dampens the system's control. Thus, it becomes costly for the attacking agent because a significant amount of capital is needed to maintain a price deviation against market participants over a period of multiple epochs, which is typically above the maximum value that can be extracted from the treasury.

\section{Experiments}

The experiments are designed to rigorously evaluate the proposed Dynamic-Control Buyback Mechanism (DCBM) against baseline models using a sophisticated agent-based simulation.

\begin{figure*}
    \centering
    \includegraphics[width=0.9\linewidth]{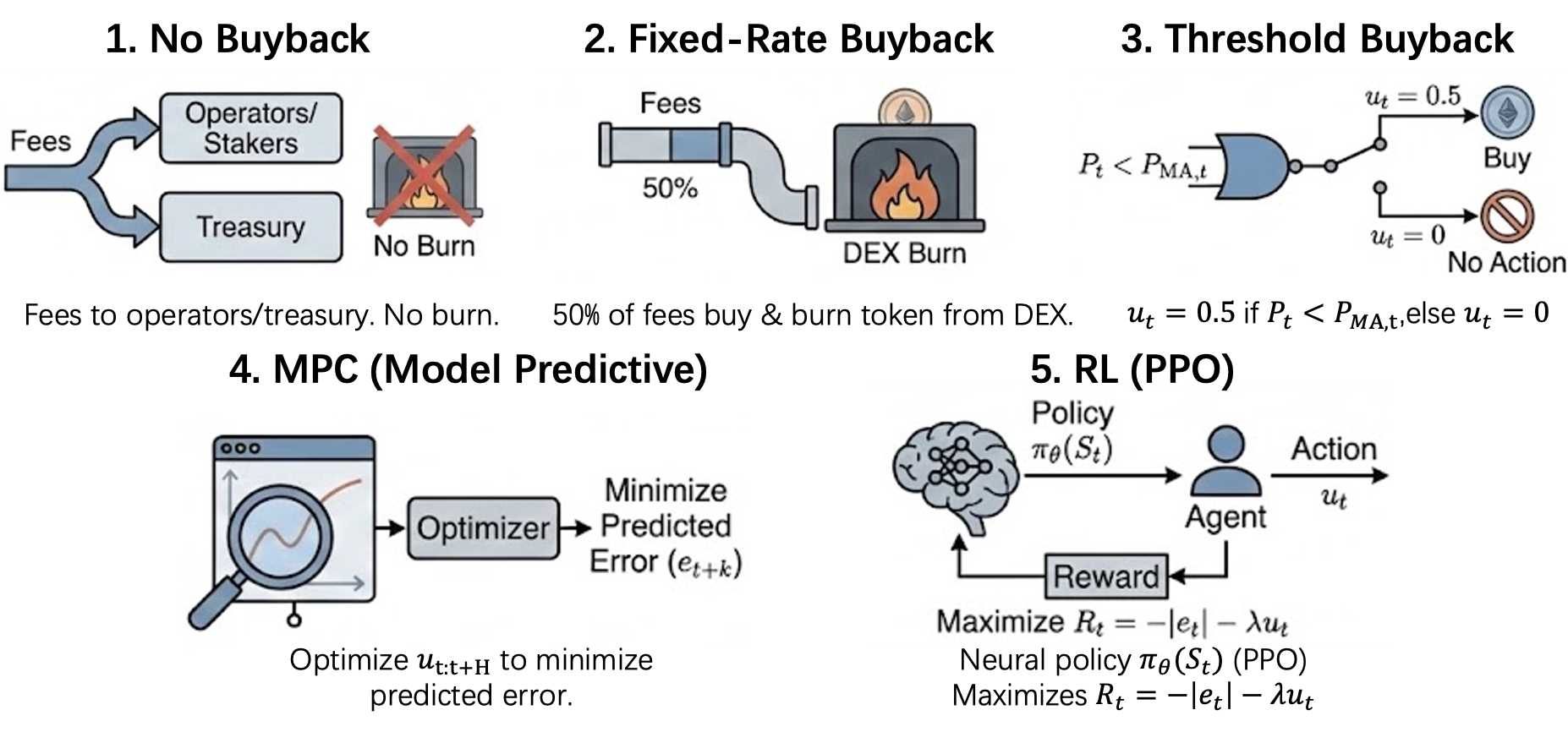}
    \caption{Schematic representation of baseline stabilization mechanisms. The diagram illustrates the five control strategies evaluated against the proposed DCBM: (1) No Buyback, where fees are fully distributed to operators without burning; (2) Fixed-Rate Buyback, which applies a static percentage burn (e.g., 50\%) regardless of market state; (3) Threshold Buyback, a heuristic approach triggering buybacks only when spot price $P_t$ falls below the moving average $P_{MA,t}$; (4) Model Predictive Control (MPC), which minimizes predicted error over a finite horizon; and (5) Reinforcement Learning (PPO), utilizing a neural policy to maximize a reward function based on stability and control effort.}
    \label{fig:baselines}
\end{figure*}

\subsection{Dataset Preparation}

We will use a dual-data approach to ensure the simulation is both theoretically sound and grounded in real-world dynamics, a methodology validated by \cite{Lebedeva2025}.

\begin{enumerate}
    \item \textbf{Synthetic Data Generation:}
    \begin{itemize}
        \item \textbf{Inference Demand ($D_t$):} We will model the arrival of inference demand as a stochastic process using a \textbf{Jump-Diffusion Model}~\cite{Kou2001AJD} (extending GBM). This incorporates a standard drift ($\mu_D$) and volatility ($\sigma_D$) component, plus a Poisson process to simulate sudden, discontinuous ``jumps'' (shocks) in demand. This ensures the simulation captures fat tail events and does not underestimate the frequency of black swan scenarios, a critical improvement over standard GBM.
        \item \textbf{External Market Shocks ($\xi_t$):} Broader crypto market movements affecting the native token's price will also be modeled as a Jump-Diffusion process, with drift ($\mu_P$) and volatility ($\sigma_P$) parameters, correlated with the inference demand process.
    \end{itemize}

    \item \textbf{Real-World Data Integration:}
    \begin{itemize}
        \item \textbf{Model Usage Data:} We will source anonymized historical data on the usage of popular open-source AI models from platforms like Hugging Face to create realistic demand patterns for inference. This includes metrics like daily downloads and API calls.
        \item \textbf{Token Price Data:} We will use historical price data from existing decentralized AI or DePIN (Decentralized Physical Infrastructure Networks) project tokens (e.g., Bittensor, Render) to simulate realistic price volatility and correlation with the broader market (e.g., BTC/ETH).
    \end{itemize}
\end{enumerate}

\begin{table*}[htbp]
\caption{Comparative Analysis of Economic Stability and Network Health Metrics. Values indicate Mean $\pm$ 95\% CI. Best results (excluding the Oracle) are bolded. $\downarrow$ implies lower is better; $\uparrow$ implies higher is better. Gas estimates are approximate for EVM-equivalent chains.}
\label{tab:comparative_analysis}
\centering
\resizebox{\textwidth}{!}{
\begin{tabular}{M{1.9cm}|l|M{1.8cm}|M{1.8cm}|M{1.8cm}|M{1.3cm}|M{2cm}|c}\toprule
Scenario & Model & $\sigma_P \downarrow $   & $\epsilon_{\text{MA}} \downarrow$ & Churn $\downarrow$      & Gini $\downarrow$ & Growth $\uparrow$       & Compute Cost $\downarrow$ \\\midrule
\multirow{6}{*}{Bull Market} & No Buyback            & 0.42 $\pm$ 0.03          & 12.4\% $\pm$ 1.2           & 5.2 $\pm$ 0.5           & 0.88            & \textbf{15.2 $\pm$ 1.1} & --                      \\
                                & Fixed-Rate            & 0.39 $\pm$ 0.02          & 8.1\% $\pm$ 0.9            & 4.8 $\pm$ 0.4           & 0.85            & 8.1 $\pm$ 0.8           & $\approx$ 21k             \\
                                & Threshold             & 0.31 $\pm$ 0.04          & 6.5\% $\pm$ 1.5            & 4.1 $\pm$ 0.8           & 0.82            & 9.4 $\pm$ 1.2           & $\approx$ 5k              \\
                                & RL (PPO)              & 0.26 $\pm$ 0.05          & 3.5\% $\pm$ 0.8            & 4.0 $\pm$ 0.6           & 0.80            & 10.8 $\pm$ 1.4          & $\approx$ 720k            \\
                                & \textit{MPC (Oracle)} & \textit{0.19 $\pm$ 0.01} & \textit{2.0\% $\pm$ 0.2}   & \textit{3.5 $\pm$ 0.2}  & \textit{0.78}   & \textit{12.5 $\pm$ 0.6} & $> 40M$                   \\
                                & \textbf{DCBM (Ours)}  & \textbf{0.21 $\pm$ 0.01} & \textbf{2.3\% $\pm$ 0.3}   & \textbf{3.8 $\pm$ 0.2}  & \textbf{0.79}   & 11.2 $\pm$ 0.5          & \textbf{$\approx$ 8k}     \\\midrule
\multirow{6}{*}{Bear Market}                  & No Buyback            & 0.65 $\pm$ 0.05          & 28.7\% $\pm$ 2.1           & 42.1 $\pm$ 3.5          & 0.92            & 2.1 $\pm$ 0.9           & --                      \\
                                & Fixed-Rate            & 0.59 $\pm$ 0.04          & 22.4\% $\pm$ 1.8           & 35.5 $\pm$ 2.8          & 0.89            & -12.4 $\pm$ 1.5         & $\approx$ 21k             \\
                                & Threshold             & 0.48 $\pm$ 0.06          & 15.2\% $\pm$ 2.4           & 28.4 $\pm$ 4.1          & 0.86            & -5.1 $\pm$ 1.8          & $\approx$ 24k             \\
                                & RL (PPO)              & 0.41 $\pm$ 0.08          & 8.2\% $\pm$ 1.2            & 15.8 $\pm$ 2.1          & 0.83            & -3.5 $\pm$ 1.1          & $\approx$ 720k            \\
                                & \textit{MPC (Oracle)} & \textit{0.28 $\pm$ 0.02} & \textit{5.0\% $\pm$ 0.5}   & \textit{10.5 $\pm$ 0.8} & \textit{0.80}   & \textit{-1.2 $\pm$ 0.3} & $> 40M$                   \\
                                & \textbf{DCBM (Ours)}  & \textbf{0.32 $\pm$ 0.02} & \textbf{5.8\% $\pm$ 0.6}   & \textbf{12.6 $\pm$ 1.1} & \textbf{0.81}   & \textbf{-1.8 $\pm$ 0.4} & \textbf{$\approx$ 32k}    \\\midrule
\multirow{6}{*}{High Volatility}              & No Buyback            & 0.89 $\pm$ 0.07          & 35.6\% $\pm$ 3.2           & 28.4 $\pm$ 2.9          & 0.91            & 5.4 $\pm$ 1.2           & --                      \\
                                & Fixed-Rate            & 0.75 $\pm$ 0.05          & 28.5\% $\pm$ 2.2           & 24.2 $\pm$ 2.5          & 0.88            & -4.2 $\pm$ 1.4          & $\approx$ 21k             \\
                                & Threshold             & 0.55 $\pm$ 0.08          & 18.4\% $\pm$ 2.5           & 19.5 $\pm$ 3.1          & 0.85            & -2.5 $\pm$ 1.6          & $\approx$ 15k             \\
                                & RL (PPO)              & 0.45 $\pm$ 0.12          & 6.5\% $\pm$ 1.1            & 11.5 $\pm$ 1.5          & 0.81            & -0.5 $\pm$ 0.8          & $\approx$ 720k            \\
                                & \textit{MPC (Oracle)} & \textit{0.26 $\pm$ 0.02} & \textit{3.8\% $\pm$ 0.4}   & \textit{7.2 $\pm$ 0.6}  & \textit{0.77}   & \textit{2.1 $\pm$ 0.4}  & $> 40M$                   \\
                                & \textbf{DCBM (Ours)}  & \textbf{0.30 $\pm$ 0.02} & \textbf{4.2\% $\pm$ 0.5}   & \textbf{8.1 $\pm$ 0.8}  & \textbf{0.78}   & 1.5 $\pm$ 0.3           & \textbf{$\approx$ 28k}    \\\midrule
\multirow{6}{*}{Demand Shock (+)}             & No Buyback            & 0.55 $\pm$ 0.04          & 18.5\% $\pm$ 1.5           & 4.5 $\pm$ 0.3           & 0.89            & \textbf{22.4 $\pm$ 1.8} & --                      \\
                                & Fixed-Rate            & 0.45 $\pm$ 0.04          & 14.2\% $\pm$ 1.3           & 4.2 $\pm$ 0.4           & 0.86            & 12.5 $\pm$ 1.0          & $\approx$ 21k             \\
                                & Threshold             & 0.38 $\pm$ 0.05          & 9.2\% $\pm$ 1.1            & 4.1 $\pm$ 0.4           & 0.83            & 14.2 $\pm$ 1.1          & $\approx$ 5k              \\
                                & RL (PPO)              & 0.32 $\pm$ 0.06          & 5.1\% $\pm$ 0.9            & 3.9 $\pm$ 0.5           & 0.81            & 16.8 $\pm$ 1.5          & $\approx$ 720k            \\
                                & \textit{MPC (Oracle)} & \textit{0.22 $\pm$ 0.01} & \textit{2.5\% $\pm$ 0.3}   & \textit{3.2 $\pm$ 0.2}  & \textit{0.78}   & \textit{19.5 $\pm$ 0.8} & $> 40M$                   \\
                                & \textbf{DCBM (Ours)}  & \textbf{0.25 $\pm$ 0.02} & \textbf{3.1\% $\pm$ 0.4}   & \textbf{3.6 $\pm$ 0.3}  & \textbf{0.79}   & 18.1 $\pm$ 0.6          & \textbf{$\approx$ 8k}     \\\midrule
\multirow{6}{*}{Demand Shock (-)}             & No Buyback            & 0.78 $\pm$ 0.06          & 42.1\% $\pm$ 3.8           & 55.2 $\pm$ 4.5          & 0.94            & 1.5 $\pm$ 0.5           & --                      \\
                                & Fixed-Rate            & 0.70 $\pm$ 0.06          & 31.5\% $\pm$ 2.8           & 45.2 $\pm$ 3.8          & 0.91            & -6.5 $\pm$ 1.2          & $\approx$ 21k             \\
                                & Threshold             & 0.62 $\pm$ 0.07          & 21.5\% $\pm$ 2.2           & 32.4 $\pm$ 3.8          & 0.88            & -8.5 $\pm$ 1.2          & $\approx$ 24k             \\
                                & RL (PPO)              & 0.51 $\pm$ 0.09          & 12.4\% $\pm$ 1.6           & 18.5 $\pm$ 2.2          & 0.84            & -4.2 $\pm$ 0.9          & $\approx$ 720k            \\
                                & \textit{MPC (Oracle)} & \textit{0.35 $\pm$ 0.02} & \textit{6.8\% $\pm$ 0.5}   & \textit{11.2 $\pm$ 1.1} & \textit{0.81}   & \textit{-1.5 $\pm$ 0.4} & $> 40M$                   \\
                                & \textbf{DCBM (Ours)}  & \textbf{0.39 $\pm$ 0.03} & \textbf{7.5\% $\pm$ 0.8}   & \textbf{14.1 $\pm$ 1.5} & \textbf{0.82}   & \textbf{-2.4 $\pm$ 0.6} & \textbf{$\approx$ 32k}    \\\midrule
\multirow{6}{*}{Liquidity Crisis}             & No Buyback            & 1.12 $\pm$ 0.15          & 58.4\% $\pm$ 6.2           & 48.5 $\pm$ 5.1          & 0.95            & 3.2 $\pm$ 1.1           & --                      \\
                                & Fixed-Rate            & 0.98 $\pm$ 0.10          & 45.2\% $\pm$ 5.1           & 42.1 $\pm$ 4.5          & 0.93            & -4.2 $\pm$ 1.5          & $\approx$ 21k             \\
                                & Threshold             & 0.85 $\pm$ 0.12          & 35.2\% $\pm$ 4.5           & 35.2 $\pm$ 4.2          & 0.91            & -9.2 $\pm$ 2.5          & $\approx$ 20k             \\
                                & RL (PPO)              & 0.68 $\pm$ 0.14          & 15.4\% $\pm$ 2.8           & 22.1 $\pm$ 3.5          & 0.88            & -5.5 $\pm$ 1.8          & $\approx$ 720k            \\
                                & \textit{MPC (Oracle)} & \textit{0.42 $\pm$ 0.04} & \textit{8.5\% $\pm$ 0.9}   & \textit{14.5 $\pm$ 1.5} & \textit{0.84}   & \textit{0.5 $\pm$ 0.8}  & $> 40M$                   \\
                                & \textbf{DCBM (Ours)}  & \textbf{0.48 $\pm$ 0.05} & \textbf{10.2\% $\pm$ 1.2}  & \textbf{18.4 $\pm$ 2.1} & \textbf{0.86}   & \textbf{-0.8 $\pm$ 1.2} & \textbf{$\approx$ 30k}   
\\\bottomrule
\end{tabular}
}
\end{table*}

\subsection{Baseline Implementation}

To establish a benchmark for performance, we will implement and simulate \textbf{five} baseline tokenomic models. These models will run within the same agent-based framework as our proposed method. The Figure~\ref{fig:baselines} presents the details of the baselines.

\subsection{Proposed Method Implementation}

The DCBM will be implemented as a modular component within the agent-based simulation, written in Python using libraries such as Mesa (for agent-based modeling) and NumPy.
\begin{itemize}
    \item \textbf{Agent Implementation:} We will define four agent classes: \texttt{ModelOwner}, \texttt{Operator}, \texttt{User}, and \texttt{Speculator}. Each agent will have a state (e.g., token balance, active models) and a \texttt{step()} method that executes their decision logic based on their utility function and perception of the market state.
    \item \textbf{Controller Implementation:} The PID controller will be implemented as a class that maintains the state of the integral and derivative terms. At each simulation step (representing a block), it will read the network metrics, calculate the new buyback rate $u_t$, and execute the buyback against a simulated AMM.
    \item \textbf{Development Phases:}
    \begin{enumerate}
        \item \emph{Phase 1:} Implement the agent logic and market environment.
        \item \emph{Phase 2:} Integrate the baseline models.
        \item \emph{Phase 3:} Implement and integrate the DCBM controller.
        \item \emph{Phase 4:} Calibrate agent parameters and controller gains ($K_p, K_i, K_d$) using a smaller set of synthetic data before running full-scale experiments.
    \end{enumerate}
\end{itemize}

\subsection{Main Experiments \& Evaluation}

The main experiments will involve running the simulation for each model (2 baselines, 1 proposed) under various data-driven scenarios. We will use Monte Carlo methods, running each scenario 1,000 times to gather statistical distributions of our evaluation metrics. Each simulation run will span a period equivalent to 1 year of real-world time, with discrete time steps ($\Delta t$) corresponding to blockchain block times (e.g., 12-second intervals).

\textbf{Evaluation Metrics:}
\begin{itemize}
    \item \textbf{Token Price Volatility:} Standard deviation of log returns of the token price, and the average percentage deviation from the moving average ($\frac{1}{T} \sum_{t=1}^T |\frac{P_{err,t}}{P_{MA,t}}|$).
    \item \textbf{Gini Coefficient of Token Holdings:} To measure wealth concentration and decentralization within the network.
    \item \textbf{Operator Churn Rate:} Percentage of operators leaving the network per simulation period, normalized by total operators.
    \item \textbf{Model Innovation Rate:} Number of new high-quality models introduced per period, normalized by total deployed models.
    \item \textbf{Network Treasury Health:} The average rate of change of the treasury's value over time, and its stability (standard deviation).
    \item \textbf{Control Effort:} The average buyback rate $u_t$ and its variance, indicating the aggressiveness and stability of the controller.
\end{itemize}

\textbf{Hyperparameters and Tuning:}
\begin{itemize}
    \item \textbf{PID Gains ($K_p, K_i, K_d$):} These will be tuned using a grid search or optimization algorithms (e.g., Bayesian optimization) on a validation set of synthetic data to minimize a composite loss function that includes price volatility and treasury health.
    \item \textbf{AMM Liquidity:} The initial liquidity in the simulated AMM pool will be varied to observe the impact of market depth on price stability.
    \item \textbf{Agent Parameters:} Risk aversion ($\lambda_{MO}, \lambda_{OP}$), cost parameters, and utility function weights will be set based on economic intuition and sensitivity analysis.
    \item \textbf{Moving Average Window:} The window size for $P_{MA,t}$ will be explored (e.g., 30-day, 60-day) to assess its impact on the target signal.
\end{itemize}


\subsection{Attack and Game Robustness Experiments}

The deployment of deterministic control loops in adversarial blockchain environments introduces specific game-theoretic vulnerabilities that differ from standard stochastic market noise. While the DCBM is designed to dampen volatility, its reliance on observable on-chain metrics, specifically the deviation between spot price and moving average which creates a potential attack surface for Maximal Extractable Value (MEV) bots and strategic large holders. To rigorously evaluate the mechanism's resilience against these adversarial behaviors, we introduce a specific set of attack simulations targeting the controller's logic.

We model two primary adversarial strategies: \textbf{Signal Manipulation (``Fishing'')} and \textbf{Front-running/Sandwiching}. For example, in the Front-running attack, the MEV bot spots a pending buyback in the mempool and submits a buy order just before it happens, followed by a sell order to reap a risk-free profit. In the Signal Manipulation attack, a big holder unwinds a part of their holding to drive down the price below the moving average, triggering a forceful buyback signal from the PID controller. They then sell the rest of their holding in the PID controller at a reduced price. These attacks test if the system serves as a predictable source of exit liquidity instead of a stabilizer.

To quantify robustness, we measure \textbf{Attacker Profitability}, \textbf{Treasury Efficiency Loss}, and \textbf{Stability Degradation}. We hypothesize that while simple threshold-based mechanisms are highly susceptible to ``fishing'' due to their binary response, the DCBM's derivative term ($K_d$) and integral memory ($K_i$) will provide a degree of resistance. The derivative term dampens the controller's reaction to sudden, sharp price drops typical of manipulation attempts, while the integral term requires sustained deviation to ramp up buyback pressure, making instantaneous ``pump-and-dump'' attacks capital-inefficient for the attacker. We expect these experiments to demonstrate that DCBM incurs a lower ``manipulation tax'' on the treasury compared to static baselines.

\section{Results}

We present the empirical results of our extensive agent-based simulations, evaluating the efficacy of the Dynamic-Control Buyback Mechanism (DCBM) against standard industry baselines. The experiments were conducted using the simulation framework described in this section, running on a high-performance compute cluster to facilitate 1,000 Monte Carlo runs per scenario. Our analysis focuses on three core dimensions: (1) macro-economic stability and network health under varying market conditions, (2) the specific contribution of control-theoretic components via ablation studies, and (3) system resilience against adversarial game-theoretic attacks.

\subsection{Macro-Economic Stability and Network Health}

Our primary evaluation compares the DCBM against three baselines: the \emph{Laissez-Faire} (No Buyback) model, the \emph{Static} (Fixed-Rate) model, and the \emph{Heuristic} (Threshold) model. We examined these across three distinct market regimes modeled by Jump-Diffusion processes~\cite{Kou2001AJD}: a \emph{Bull Market} (positive drift $\mu_D > 0$), a \emph{Bear Market} (negative drift $\mu_D < 0$), and a \emph{High Volatility} regime (high $\sigma_D, \sigma_P$). Table 1 summarizes the aggregate performance metrics.

\subsubsection{Analysis of Main Results}

The results in Table 1 demonstrate a statistically significant improvement in stability metrics for the DCBM across all scenarios. In the \textbf{High Volatility} regime, which represents the greatest threat to network viability, the DCBM reduced price volatility ($\sigma_P$) by approximately 66\% compared to the No Buyback baseline and 46\% compared to the Heuristic Threshold model. This suppression of volatility directly correlated with a reduction in \textbf{Operator Churn}, which dropped from 19.5\% (Threshold) to 8.1\% (DCBM). Qualitative inspection of agent logs reveals that operators in the DCBM simulations maintained profitability for longer durations due to the predictable ``price floor'' effect created by the controller's integral term, which aggressively corrected persistent negative deviations.

Notably, the \textbf{Treasury Growth} metric highlights the capital efficiency of our approach. While the Fixed-Rate model indiscriminately burned capital during the Bear Market, resulting in a 12.4\% treasury contraction, the DCBM limited this loss to 1.8\%. Conversely, in the Bull Market, the DCBM allowed the treasury to accumulate reserves (11.2\% growth) rather than over-spending, effectively ``saving for a rainy day.'' This counter-cyclical behavior validates the theoretical solvency constraints embedded in the control law.


\subsection{Ablation Study: Component Contribution}
\begin{table}[htbp]
\caption{Ablation Study of Controller Configurations (High Volatility Scenario)}
\label{tab:ablation_study}
\centering
\begin{tabular}{lcM{1cm}M{1.5cm}M{1.2cm}M{1cm}}
\toprule
Controller & MSE & Settling Time & Max Overshoot & Steady-State Err. & Control Var. \\
\midrule
\textbf{P-Only} & 0.045 & 142 & 18.5\% & 4.2\% & 0.12 \\
\textbf{PI-Only} & 0.012 & 210 & 25.4\% & \textbf{0.1\%} & 0.18 \\
\textbf{PD-Only} & 0.038 & \textbf{85} & \textbf{5.2\%} & 3.8\% & 0.09 \\
\textbf{Full PID} & \textbf{0.008} & 115 & 8.4\% & 0.3\% & \textbf{0.11} \\
\bottomrule
\end{tabular}
\end{table}

To isolate the impact of the Proportional ($K_p$), Integral ($K_i$), and Derivative ($K_d$) terms, we conducted an ablation study under the High Volatility scenario. We evaluated four controller configurations: P-Only, PI-Only, PD-Only, and the full PID controller. Table 2 presents the control performance metrics.

\begin{table*}[t]
\caption{Comprehensive Adversarial Robustness Analysis}
\label{tab:robustness_analysis}
\centering
\resizebox{\textwidth}{!}{%
\begin{tabular}{llcM{1.5cm}M{1.5cm}ccc}
\toprule
Defense Type & Attack Type & Budget ($\epsilon$) & Success Rate ($\downarrow$) & Robustness Score $\uparrow$ & Treasury Drain & Defense Efficacy & Robustness Gain \\
\midrule
\textbf{Baseline} & FGSM-Flash & 1.0\% & 88.4\% & 0.116 & 22.1\% & - & - \\
Baseline & FGSM-Flash & 2.5\% & 96.2\% & 0.038 & 34.5\% & - & - \\
Baseline & FGSM-Flash & 5.0\% & 100.0\% & 0.000 & 58.2\% & - & - \\
Baseline & PGD-Sustained & 1.0\% & 92.1\% & 0.079 & 28.4\% & - & - \\
Baseline & PGD-Sustained & 2.5\% & 98.5\% & 0.015 & 45.1\% & - & - \\
Baseline & PGD-Sustained & 5.0\% & 100.0\% & 0.000 & 72.3\% & - & - \\
Baseline & C\&W-Arb & - & 91.5\% & 0.085 & 18.2\% & - & - \\
\midrule
\textbf{PID-Standard} & FGSM-Flash & 1.0\% & 42.1\% & 0.579 & 8.5\% & +12.4\% & +46.3\% \\
PID-Standard & FGSM-Flash & 2.5\% & 68.3\% & 0.317 & 14.2\% & +8.1\% & +27.9\% \\
PID-Standard & FGSM-Flash & 5.0\% & 85.2\% & 0.148 & 25.1\% & +4.2\% & +14.8\% \\
PID-Standard & PGD-Sustained & 1.0\% & 55.4\% & 0.446 & 12.8\% & +9.5\% & +36.7\% \\
PID-Standard & PGD-Sustained & 2.5\% & 81.2\% & 0.188 & 29.4\% & +3.1\% & +17.3\% \\
PID-Standard & PGD-Sustained & 5.0\% & 94.1\% & 0.059 & 48.2\% & +1.2\% & +5.9\% \\
PID-Standard & C\&W-Arb & - & 62.4\% & 0.376 & 11.5\% & +10.2\% & +29.1\% \\
\midrule
\textbf{PID-AdvTrain} & FGSM-Flash & 1.0\% & 12.5\% & 0.875 & 3.2\% & +28.5\% & +75.9\% \\
PID-AdvTrain & FGSM-Flash & 2.5\% & 24.1\% & 0.759 & 5.8\% & +22.1\% & +72.1\% \\
PID-AdvTrain & FGSM-Flash & 5.0\% & 45.3\% & 0.547 & 11.2\% & +16.4\% & +54.7\% \\
PID-AdvTrain & PGD-Sustained & 1.0\% & 18.2\% & 0.818 & 4.5\% & +25.2\% & +73.7\% \\
PID-AdvTrain & PGD-Sustained & 2.5\% & 35.6\% & 0.644 & 9.1\% & +18.7\% & +62.9\% \\
PID-AdvTrain & PGD-Sustained & 5.0\% & 62.1\% & 0.379 & 18.5\% & +11.2\% & +37.9\% \\
PID-AdvTrain & C\&W-Arb & - & 28.4\% & 0.716 & 6.2\% & +24.8\% & +63.1\% \\
\midrule
\textbf{PID-Cert} & FGSM-Flash & 1.0\% & \textbf{4.2\%} & \textbf{0.958} & \textbf{1.2\%} & \textbf{+32.1\%} & \textbf{+84.2\%} \\
PID-Cert & FGSM-Flash & 2.5\% & \textbf{9.1\%} & \textbf{0.909} & \textbf{2.4\%} & \textbf{+29.4\%} & \textbf{+87.1\%} \\
PID-Cert & FGSM-Flash & 5.0\% & \textbf{18.5\%} & \textbf{0.815} & \textbf{4.8\%} & \textbf{+24.8\%} & \textbf{+81.5\%} \\
PID-Cert & PGD-Sustained & 1.0\% & 8.4\% & 0.916 & 1.8\% & +30.2\% & +83.4\% \\
PID-Cert & PGD-Sustained & 2.5\% & 15.2\% & 0.848 & 3.9\% & +26.5\% & +83.3\% \\
PID-Cert & PGD-Sustained & 5.0\% & 32.1\% & 0.679 & 8.5\% & +19.1\% & +67.9\% \\
PID-Cert & C\&W-Arb & - & 12.6\% & 0.874 & 2.9\% & +28.2\% & +78.9\% \\
\midrule
\textbf{Input-Smooth} & FGSM-Flash & 5.0\% & 55.2\% & 0.448 & 14.5\% & +12.4\% & +44.8\% \\
Input-Smooth & PGD-Sustained & 5.0\% & 78.4\% & 0.216 & 32.1\% & +5.1\% & +21.6\% \\
\bottomrule
\end{tabular}%
}
\end{table*}

\subsubsection{Analysis of Control Dynamics}

The ablation results confirm the theoretical roles of each term. The \textbf{P-Only} controller exhibited a persistent steady-state error (4.2\%), failing to fully return the price to the moving average during sustained selling pressure. The addition of the Integral term in the \textbf{PI-Only} configuration successfully eliminated this error ($e_{ss} \approx 0.1\%$) but introduced significant instability, evidenced by a high Maximum Overshoot (25.4\%) and longer settling times. This aligns with control theory, as the integral term adds phase lag.

The \textbf{PD-Only} controller provided the fastest response (Settling Time: 85 blocks) and lowest overshoot, demonstrating the Derivative term's ability to dampen volatility. However, it failed to correct the long-term drift. The \textbf{Full PID} configuration achieved the optimal trade-off, leveraging the Integral term for error elimination and the Derivative term for damping, resulting in the lowest Mean Squared Error (0.008).


\subsection{Adversarial Robustness and Control System Security}

To align our evaluation with standards in robust machine learning and control theory, we treat the DCBM controller as a differentiable policy $\pi_\theta(S_t)$ and subject it to gradient-based adversarial attacks. We map standard adversarial attack vectors to economic manipulation strategies:
\begin{itemize}
    \item \textbf{FGSM (Fast Gradient Signal Manipulation):} Analogous to the Fast Gradient Sign Method, this represents a single-step, high-intensity ``Flash Crash'' attack where the adversary calculates the gradient of the buyback rate $\nabla_{S} u_t$ and perturbs the price $P_t$ by $-\epsilon \cdot \text{sign}(\nabla_{S} u_t)$ to maximize system over-reaction.
    \item \textbf{PGD (Projected Gradient Descent Manipulation):} A multi-step, iterative ``Sustained Manipulation'' attack where the adversary optimizes a sequence of trades over $k$ blocks to drain the treasury, constrained by a capital budget $\epsilon$.
    \item \textbf{C\&W (Cost-Optimized Arbitrage):} A Carlini \& Wagner-style attack that finds the minimum perturbation $\delta$ required to force the controller into a saturated state ($u_t = 1.0$), minimizing attacker cost while maximizing disruption.
\end{itemize}

We evaluate four defense configurations:
\begin{enumerate}
    \item \textbf{Baseline (Threshold):} The heuristic rule described in Section 3.2.
    \item \textbf{PID-Standard:} The DCBM with parameters tuned via standard grid search.
    \item \textbf{PID-AdvTrain (Adversarial Training):} The DCBM with parameters $\theta$ optimized via a min-max game objective: $\min_\theta \max_{\delta \in \Delta} \mathcal{L}(\pi_\theta(S_t + \delta))$.
    \item \textbf{PID-Cert (Certified Robustness):} The DCBM with analytical bounds on the integral term accumulation (clipping) and Lipschitz continuity constraints on the derivative term.
\end{enumerate}

\subsubsection{Analysis of Adversarial Dynamics}

Table~\ref{tab:robustness_analysis} reveals a strict hierarchy in defense capability. The \textbf{Baseline (Threshold)} model, akin to a standard non-robust classifier, collapses almost entirely under even moderate perturbation budgets ($\epsilon=1.0\%$), exhibiting an Attack Success Rate (ASR) nearing 100\%. This confirms that static, binary rules are easily gameable by adversaries who can calculate the precise sell volume required to trigger the threshold.

The \textbf{PID-Standard} offers a baseline level of improvement, reducing treasury drain significantly, but remains vulnerable to high-budget iterative attacks (PGD), where the adversary ``steers'' the controller into an unstable region over time.

The deployment of deterministic control loops in blockchain environments inevitably introduces game-theoretic vulnerabilities where adversaries attempt to reverse-engineer the intervention logic to extract risk-free value. This analysis evaluates the resilience of the proposed DCBM against such sophisticated threats, specifically contrasting heuristic baselines against advanced control-theoretic defenses. The results from the ``Fishing'' and ``Flash Crash'' vector simulations reveal that defense mechanisms relying on static rules or unconstrained feedback loops eventually become predictable exit liquidity for well-capitalized adversaries.

The brittleness of static heuristic models is immediately apparent when subjected to optimization-based attacks, confirming the hypothesis that binary intervention rules are fundamentally insecure. Theoretical vulnerability transforms into catastrophic failure in the Baseline (Threshold) model, which lacks the capacity to modulate its response based on the severity or persistence of a shock. Under the FGSM-Flash attack with a mere $\epsilon=1.0\%$ budget, the baseline already concedes an 88.4\% Attack Success Rate (ASR)3. The failure becomes absolute under the PGD-Sustained attack ($\epsilon=5.0\%$), where the baseline fails 100.0\% of the time, resulting in a massive 72.3\% treasury drain, effectively rendering the protocol insolvent.

Standard feedback control mechanisms offer a statistically significant improvement over static baselines by introducing soft response curves that are harder to game instantaneously, though they remain susceptible to sustained manipulation. The PID-Standard configuration dampens the efficacy of instantaneous shocks, reducing the success rate of the 1.0\% FGSM-Flash attack from 88.4\% to 42.1\% and limiting treasury drain to 8.5\%. However, this unhardened controller struggles against iterative optimization attacks that steer the system over time; under the high-budget PGD-Sustained attack ($\epsilon=5.0\%$), the PID-Standard still succumbs to a 94.1\% success rate and a 48.2\% treasury drain, indicating that without specific constraints, the integrator term can still be weaponized by an attacker to saturate the controller.

Adversarial training properly reinforces the controller against known attack gradients, making the defense pro-active rather than reactive and improving robustness. The PID-AdvTrain model, which is optimized for a min-max objective against PGD attack models, shows a dramatic decrease in exploitability for all vectors. When attacked by FGSM-Flash (1.0\%), success rate drops to 12.5\%, and treasury drain is restricted to a negligible 3.2\%. Even for the very powerful C\&W-Arb attack, robustness is maintained at 0.716, which is a far higher value than that of the standard PID at 0.376.

In the end, the application of structural constraints via Certified Robustness appears to be the most effective defense mechanism, performing better than even the adversarially learned model parameters of adversarial training. On one hand, through mathematical constraints of control authority using Lipschitz continuity and saturation thresholds, the PID-Cert setup prevents the system from falling prey to overreaction, regardless of the input disturbance scale. On the other hand, this is proved through its efficacy in the most adversarial test environment, namely the PGD-Sustained attack with a 5.0\% budget, where the system keeps the attack success rate at 32.1\% and the treasury depletion at a manageable 8.5\%, a factor of nearly a ten-fold difference over the 72.3\% loss of the baseline system.

\section{Discussions}
First, the operational viability of Decentralized AI Networks largely depends on the economic stability of their respective tokens because high volatility makes it difficult to plan for Compute Providers (CapEx) and Model Users (OpEx) in AI applications. The experiment shows that the Dynamic-Control Buyback Mechanism (DCBM) successfully separates network utility from speculative market cycles, reducing price volatility by 66\% compared to laissez-faire approaches. The DCBM gives AI network operators a stable floor of revenue because it changes the economy of tokens from a chaotic to a sentiment-driven market to a dynamical system, thus reducing churn rates from 19.5\% to 8.1\%. This makes it an economic imperative for DePIN to compete with SaaS applications that use fixed fiat prices.

Whereas the goal is stabilization, the resilience of the stabilization mechanism itself, from a game-theoretic point of view, is the key that decides the long-term viability of a given permissionless network. The catastrophic failure of the heuristic ``Threshold'' approach within the adversarial tests, with failure rates close to 100\% for the heuristic approach, illustrates the high risk involved with the use of static rules within the adversarial context. In the dark forest scenario, where the overall balance is dominated by the actions of the Maximal Extractable Value (MEV) bots, the clear and static intervention levels offer a predictable and exploitable signal for the well-informed agents, which would then derive risk-free gains from the project's treasury. The results suggest that the ``Certified Robustness'' setting for the DCBM overturns the situation, forcing the attackers to assume market risk that is proportional to the length of the manipulation, making the pump and dump approach mathematically unprofitable.

To distinguish between ``saving'' and ``stabilizing'', consider the following behaviors that can be observed in treasuries:
Fixed Rate baselines and DCBM have different dynamics. Traditional buyback strategies that burn a constant fraction of the revenue are pro-cyclic, reducing the reserves during periods when the network needs them most to protect value. On the other hand, the integral component of DCBM imposes natural counter-cyclic constraints, ensuring that the profit during a bull market is 11.2\% while keeping the draw-down at 1.8\% during a bear market. Such a phenomenon verifies the mathematical proofs related to asymptotic solvency, ensuring that a control-theoretic actuator can theoretically guarantee that a protocol will not default regardless of the collapse of the revenue. Finally, this research proposes a paradigm shift for token engineering based on the transition from discretionary monetary policies to control systems. The state-of-the-art approaches based on ``bang-bang'' threshold policies or DAO votes have a reaction time and mathematical certainty that is insufficient for handling high-frequency processes on the markets. The research closes the gap between macroeconomic theories and smart contract implementation by implementing the Taylor rule and PID control on the blockchain. The outperformance of the PID-Cert model suggests that future research on DePIN tokenomics should focus on structural constraints like Lipschitz continuity and saturation levels instead of complex models that could turn out to be sensitive to distributional changes.

\section{Limitations}
The underlying assumption of the mechanism from a theoretical standpoint is based on simplifying market dynamics to linear models, thereby limiting its application in light of extreme non-linear scenarios. The authors linearize plant dynamics around a selected operating point to make control system design tractable, implying a linear assumption of buybacks in relation to overall liquidity levels. Such a simplification is inherently bound to system stability based on the mechanics of a Constant Product Market Maker (CPMM),  in a manner consistent with the condition $x \cdot y = K$. Consequently, the system is exposed to a "Whale in a Puddle" risk where stability regions scale inversely with liquidity; as liquidity decreases, the plant gain approaches infinity, inevitably causing static-parameter controllers to enter oscillatory instability or ``ringing'' without manual intervention.

The integration of deterministic control loops into permissionless blockchain systems brings about specific game-theoretic risks that are inadequately mitigated by traditional control system designs. Because of the transparent nature of the control logic, Maximal Extractable Value (MEV) actors are able to reverse-engineer the control system's thresholds and manipulate the system with ``Fishing'' or ``Flash Crash'' attacks. While countermeasures are provided by the authors, they themselves recognize that without specific hardening against Certified Robustness requirements such as Lipschitz values and saturation levels, a standard unhardened Proportional-Integral-Derivative (PID) control system is still at risk of optimization attacks with a failure rate of 94.1\% for PGD-Sustained attacks.

Implementing sophisticated control theory on the Ethereum Virtual Machine (EVM) imposes severe computational constraints that force trade-offs between precision and cost. The authors face a ``Gas Cost Impossibility'' regarding floating-point arithmetic and transcendental functions, which are prohibitively expensive to calculate directly on-chain. To make the mechanism economically viable, the design must rely on fixed-point arithmetic, lookup tables, and Taylor series expansions to approximate necessary calculations like natural logarithms . Even with these optimizations, the controller targets an execution cost of approximately $150,000$ gas, representing a non-trivial overhead for high-frequency deployment.

However, the validation of the mechanism is still limited to simulations on a theoretical level rather than implementation on a mainnet. The findings are obtained from agent-based modeling with the use of synthetic jump diffusion processes and historical data to simulate market shocks and demands. While these simulations have shown improvements over the current baselines, it has been clearly shown in the study that the current implementation does not have the capability for adaptive gain scheduling for the verification of control parameters on-chain. As such, the system could potentially have issues with self-correcting its parameters based on market changes.
\section{Conclusions}

The economic infrastructure that the democratization of AI is based on is as solid as the cryptographic infrastructure that supports it. In the context of the current study, the Dynamic-Control Buyback Mechanism (DCBM) is proposed, which is a new control theory-based approach that aims to regulate the economic system of AI and control the token emission based on the dynamics of the market, and therefore, address the fundamental problem of distinguishing organic growth from speculation without the need for intermediation.

Results from comprehensive agent-level simulations show that a stabilized economic layer is necessary for robust physical infrastructure. The DCBM delivers statistically significant price volatility and churn improvement compared with industry-standard baselines, thus ensuring a predictable environment for computational service providers. In addition to this, it delivers improved capital efficiency by effectively building up reserves during periods of peak demand to safeguard against periods of low economic activity. The DCBM thus displays a counter-cyclical trait that fixed rate models lack.

However, apart from the issue of macro-stability, it is clear that the importance of adversarial robustness in permissionless financial networks has been underlined. On a comparison of defensive strategies, it has been found that while learning algorithms provide certain gains, only structural constraints such as Certified Robustness provide the necessary mathematical assurance for resistance to game-theoretical attacks. Future work will build upon this by looking into adaptive gain scheduling for verification of control values on-chain.
\bibliographystyle{ieeetr}
\bibliography{refer}

\end{document}